\documentclass{IEEEtran}
\usepackage{amsmath}
\usepackage{amsmath}
\usepackage{amsmath}
\usepackage{amsfonts}
\usepackage{amsmath}
\usepackage{graphicx}
\usepackage{color}
\usepackage{multicol}
\usepackage{amssymb}
\begin{document}

\title{Closed-Form Expressions for Secrecy Capacity over Correlated Rayleigh Fading Channels}
\IEEEoverridecommandlockouts
\author{Xiaojun Sun,
        Chunming Zhao,~\IEEEmembership{Member\ ~IEEE,}
         and Ming Jiang,~\IEEEmembership{Member\ ~IEEE,}
\IEEEcompsocitemizethanks{ \IEEEcompsocthanksitem National Mobile
Communications Research Laboratory, Southeast University, Nanjing
210096, CHINA. E-mail: \{sunxiaojun, cmzhao,
jiang\_ming\}@seu.edu.cn \protect\\ }
}

\maketitle

\begin{abstract}
We investigate the secure communications over correlated wiretap
Rayleigh fading channels assuming the full channel state information
(CSI) available. Based on the information theoretic formulation, we
derive closed-form expressions for the average secrecy capacity and
the outage probability. Simulation results confirm our analytical
expressions.
\end{abstract}


\section{Introduction}
Because wireless communications are susceptible to eavesdropping,
traditional security mechanisms mainly rely on cryptographic
protocols. Recently, potential benefits of deriving secure
information from physical layer have been reported in [1]. In [2],
the information-theoretic secrecy capacity was introduced by using
the physical properties of channels.

The basic principle of information-theoretic security has been
widely accepted as the strictest notion of security, which
guarantees that the sent massages can not be decoded by a malicious
eavesdropper [1]. Wyner introduced wiretap channel model to evaluate
secure transmissions at the physical layer \cite{IEEEhowto:2}, where
Alice transmits confidential data to Bob and Eve eavesdrops the
data. Csiszar \emph{et. al. } and Leung-Yan-Cheong \emph{et. al. }
generalized it to broadcast channels and basic Gaussian channels,
respectively in \cite{IEEEhowto:3} and \cite{IEEEhowto:4}. Wei Kang
\emph{et. al. } studied secure communications over two-user
semi-deterministic broadcast channels \cite{IEEEhowto:5}. The
secrecy capacity is defined as the difference between the main
channel capacity (Alice to Bob) and the eavesdropping's channel
capacity (Alice to Eve) \cite{IEEEhowto:4}. Barros \emph{et. al. }
and Gopala \emph{et. al. } generalized this Gaussian wiretap channel
model to wireless quasi-static fading channels
\cite{IEEEhowto:6}-\cite{IEEEhowto:8}. The secure MIMO systems are
studied in \cite{IEEEhowto:9}-\cite{IEEEhowto:10}. Motivated by
emerging wireless applications, there is a growing interest in
exploiting the benefits of relay and cooperative strategies in order
to guarantee secure transmissions
\cite{IEEEhowto:11}-\cite{IEEEhowto:13}.

In this paper, we consider the secure communications within Wyner's
correlated wiretap channel by building on the detailed technique in
[7]. Similar work studied in [14] gives the limiting value of the
average secrecy capacity, which only converges into the secrecy
capacity at the high signal-to-noise ratio (SNR). We derive the
closed-form expressions of the average secure communication capacity
and the outage probability  under the assumption of the full channel
state information (CSI) available. Simulation results verify our
analytical expressions.

\section{System Model}
\begin{figure}
\centering
\includegraphics[width=3in, height=1.5in]{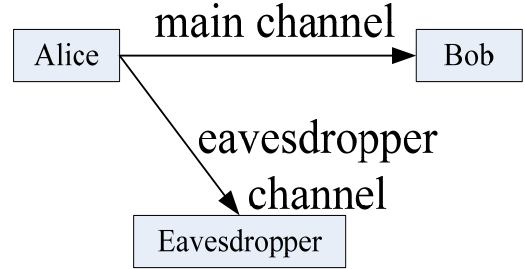}
\vspace{-0.3cm} \caption{Fading wiretap channel model.}
\end{figure}
Fig. 1 shows a fading wiretap channel model considered in this
paper. Here, the source Alice transmits confidential information to
the destination Bob. A third party (Eve) is able of eavesdropping on
the transmissions. The received signals at Bob and Eve are given by
\addtocounter{equation}{1}
\begin{align}
y = h_{sd} x + n_d
 \tag{\theequation a}\\
z = h_{se} x + n_e  \tag{\theequation b}
\end{align}
where $h_{se}$ and $h_{sd}$ are complex Gaussian random variables
(RV) with zero-mean. $N_d $ and $N_e $ denote zero-mean complex
Gaussian noise RVs with unit-variance. The instantaneous SNRs
$\alpha  = \left| {h_{sd} } \right|^2 $ and $\beta = \left| {h_{se}
} \right|^2 $ are exponentially distributed. The average value is
$\lambda _1 $ and $\lambda _2 $, respectively.

Since $\alpha$ and $\beta $ are correlated variables, the joint PDF
is expressed as [14][15]
\begin{equation}
f\left( {\alpha ,\beta } \right) = \frac{{I_0 \left( {\frac{2} {{1 -
\rho }}\sqrt {\frac{{\alpha \beta \rho }} {{\lambda _1 \lambda _2
}}} } \right)}} {{\left( {1 - \rho } \right)\lambda _1 \lambda _2
}}\exp \left( { - \frac{{\alpha /\lambda _1  + \beta /\lambda _2 }}
{{1 - \rho }}} \right)
\end{equation}
where $\rho $ is the correlation between $h_{se}$ and $h_{sd}$. $I_0
\left(  \cdot  \right)$ is the zeroth-order modified Bessel function
of the first kind. Using the infinite-series representation of  $I_0
\left( x \right)$ [16]
\begin{align*}
I_0 \left( x \right) = 1 + \sum\limits_{k = 1}^\infty {\frac{{x^{2k}
}} {{4^k \left( {k!} \right)^2 }}} \end{align*} We can rewrite the
joint PDF as
\begin{equation}
\begin{gathered}
  f\left( {\alpha ,\beta } \right) = \sum\limits_{k = 0}^\infty  {c_k } \frac{{\exp \left( { - \frac{{\alpha /\lambda _1  + \beta /\lambda _2 }}
{{1 - \rho }}} \right)}} {{\lambda _1 \lambda _2 }}\left(
{\frac{\alpha } {{\lambda _1 }}} \right)^k \left( {\frac{\beta }
{{\lambda _2 }}} \right)^k  \hfill \\
   = \sum\limits_{k = 0}^\infty  {c_k } \frac{1}
{{\lambda _1 \lambda _2 }}\exp \left( { - \frac{{\alpha /\lambda _1
+ \beta /\lambda _2 }} {{1 - \rho }}} \right)\left( {\frac{\alpha }
{{\lambda _1 }}} \right)^k \left( {\frac{\beta }
{{\lambda _2 }}} \right)^k  \hfill \\
   = \sum\limits_{k = 0}^\infty  {c_k f_k } \left( {\alpha ,\beta } \right) \hfill \\
\end{gathered}
\end{equation}
where $c_k  = \frac{{\rho ^k }} {{\left( {k!} \right)^2 \left( {1 -
\rho } \right)^{2k + 1} }}$.

In this study, we assume that Alice has access to CSI on both the
main channel and the eavesdropper's channel. For instance, Eve may
be not a covert eavesdropper, but simply another user [6][7]. Alice
wants to transmit some confidential data to Bob which Alice does not
wish Eve know. So, Alice can estimate the CSI of the eavesdropper's
channel [6][7].
\section{Secure communications over correlated Rayleigh fading channel}
Start with the technique detailed in [7], which introduces the
secrecy capacity over fading channel. A similar introduction was
presented in [8]. Recalling the results of [7] for the Rayleigh
fading wiretap channel, the secrecy capacity for one realization can
be written as
\begin{equation}
C_s \left( {\alpha ,\beta} \right) = \left\{
\begin{gathered}
  \ln \left( {1 + \alpha } \right) - \ln \left( {1 + \beta } \right)\;,\;\;if\;\;\alpha  > \beta  \hfill \\
  0\;\;\;\;\;\;\;\;\;\;\;\;\;\;\;\;\;\;\;\;\;\;\;\;\;\;\;\;,\;\;if\;\;\alpha  \leqslant \beta  \hfill \\
\end{gathered}  \right.
\end{equation}
where $\ln \left( {1 + \alpha } \right)$ is the rate of the main
channel, and $\ln \left( {1 + \beta } \right)$  denotes the rate of
the eavesdropper's channel.

\subsection{average secrecy
capacity} The average secrecy capacity over correlated channels is
derived as follows.
\newtheorem{theorem}{Theorem}
\begin{theorem}The average secrecy capacity is averaged over all channel
realizations
\begin{equation}
\begin{gathered}
  {\text{C}}_s  =  \int_0^\infty  {\int_0^\infty  {C_s \left( {\alpha ,\beta } \right)f\left( {\alpha ,\beta } \right)d\alpha d\beta } }  \hfill \\
   = \sum\limits_{k = 0}^\infty  {c_k k!\left( {1 - \rho } \right)^{k + 1} F\left( {\lambda _1 ,k,\frac{1}
{{1 - \rho }}} \right)}  \hfill \\
   - k!\sum\limits_{m = 0}^k {\frac{{\left( {\lambda _1 /\lambda _2 } \right)^m }}
{{m!}}\left( {1 - \rho } \right)^{k + 1 - m} } F\left( {\lambda _1
,k + m,\frac{{1 + \lambda _1 /\lambda _2 }}
{{1 - \rho }}} \right) \hfill \\
   - k!\sum\limits_{m = 0}^k {\frac{{\left( {\lambda _2 /\lambda _1 } \right)^m }}
{{m!}}} \left( {1 - \rho } \right)^{k + 1 - m} F\left( {\lambda _2
,k + m,\frac{{1 + \lambda _2 /\lambda _1 }}
{{1 - \rho }}} \right) \hfill \\
\end{gathered}
\end{equation}
\end{theorem}
where function $F\left( {\lambda ,k,\mu } \right)$ can be
recursively evaluated or computed by using popular symbolic software
like MATLAB. After integration by parts, we get
\begin{equation}
\begin{gathered}
  F\left( {\lambda ,k,\mu } \right) = \int_0^\infty  {\ln \left( {1 + \lambda x} \right)\exp \left( { - \mu x} \right)x^k dx}  \hfill \\
   = \frac{\lambda }
{\mu }F_k  + \frac{k}
{\mu }F\left( {\lambda ,k - 1,\mu } \right) \hfill \\
\end{gathered}
\end{equation}
where $F_k $ is defined by [16, 3.353.5]
\begin{align*}
\begin{gathered}
  F_k  = \int_0^\infty  {\frac{{x^k }}
{{1{\text{ + }}\lambda x}}} e^{ - \mu x} dx \hfill \\
   = \frac{1}
{\lambda }\left[ {\left( { - \frac{1} {\lambda }} \right)^k \exp
\left( {\frac{\mu } {\lambda }} \right){\rm E}_1 \left( {\frac{\mu }
{\lambda }} \right) + \sum\limits_{m = 1}^k {\Gamma \left( m
\right)\frac{{\left( { - \lambda } \right)^{m - k} }}
{{\mu ^m }}} } \right] \hfill \\
\end{gathered}.
\end{align*}
Function $F\left( {\lambda ,0,\mu } \right) = \frac{1} {\mu }E_1
\left( {\frac{\mu } {\lambda }} \right)\exp \left( {\frac{\mu }
{\lambda }} \right)$. $E_1 \left( x \right) = \int_1^\infty
{\frac{{e^{ - xt} }} {t}dt} $ is the exponential-integral function
[16]. $\Gamma \left( { \cdot} \right)$ is the gamma function [16].

\begin{IEEEproof}
The integral in (5) is re-expressed as
\begin{align*}
\begin{gathered}
  {\text{C}}_s = \sum\limits_{k = 0}^\infty  {c_k } \int_0^\infty  {\int_0^\alpha  {\ln \left( {1 + \alpha } \right)f_k \left( {\alpha ,\beta } \right)d\alpha d\beta } }  \hfill \\
   - \int_0^\infty  {\int_\beta ^\infty  {\ln \left( {1 + \beta } \right)f_k \left( {\alpha ,\beta } \right)d\alpha d\beta } }  \hfill \\
   = \sum\limits_{k = 0}^\infty  {c_k } \int_0^\infty  {\int_0^{\lambda _1 u/\lambda _2 } {\ln \left( {1 + \lambda _1 u} \right)} } f_k \left( {u,v} \right)dudv \hfill \\
   - \int_0^\infty  {\int_{\lambda _2 v/\lambda _1 }^\infty  {\ln \left( {1 + \lambda _2 v} \right)} } f_k \left( {u,v} \right)dudv \hfill \\
   = \sum\limits_{k = 0}^\infty  {c_k } \left( {R_k^1  - R_k^2 } \right) \hfill \\
\end{gathered}
\end{align*}
The integral $R_k^1 $ can be evaluated by[16, 3.381.1]
\begin{equation}
\begin{gathered}
  R_k^1  = \int_0^\infty  {\ln \left( {1 + \lambda _1 u} \right)\exp \left( { - \frac{u}
{{1 - \rho }}} \right)u^k du}  \hfill \\
   \times \int_0^{\lambda _1 u/\lambda _2 } {\exp \left( { - \frac{v}
{{1 - \rho }}} \right)v^k dv}  \hfill \\
   = \int_0^\infty  {\left( {1 - \rho } \right)^{k + 1} \ln \left( {1 + \lambda _1 u} \right)\exp \left( { - \frac{u}
{{1 - \rho }}} \right)u^k }  \hfill \\
   \times \gamma \left( {k + 1,\frac{{\lambda _1 u}}
{{\lambda _2 (1 - \rho )}}} \right)du \hfill \\
\end{gathered}
\end{equation}
where $\gamma \left( { \cdot , \cdot } \right)$ is the incomplete
gamma function  defined by [16]
\begin{align*}
\gamma \left( {n + 1,x} \right) = n! - n!e^{ - x} \sum\limits_{m =
0}^n {\frac{{x^m }} {{m!}}}
\end{align*}
Further with some manipulations, $R_k^1 $ can be rewritten as
\begin{equation}
\begin{gathered}
  R_k^1  = \int_0^\infty  {k!\left( {1 - \rho } \right)^{k + 1} \ln \left( {1 + \lambda _1 u} \right)\exp \left( { - \frac{u}
{{1 - \rho }}} \right)u^k du}  \hfill \\
   - k!\sum\limits_{m = 0}^k {\frac{{\left( {\lambda _1 /\lambda _2 } \right)^m }}
{{m!}}} \int_0^\infty  {\left( {1 - \rho } \right)^{k + 1 - m} \ln \left( {1 + \lambda _1 u} \right)}  \hfill \\
   \times \exp \left( { - \frac{{1 + \lambda _1 /\lambda _2 }}
{{1 - \rho }}u} \right)u^{k + m} du \hfill \\
\end{gathered}
\end{equation}
Using (6), we can evaluate the integral
\begin{equation}
\begin{gathered}
  R_k^1  = k!\left( {1 - \rho } \right)^{k + 1} F\left( {\lambda _1 ,k,\frac{1}
{{1 - \rho }}} \right) \hfill \\
   - k!\sum\limits_{m = 0}^k {\frac{{\left( {\lambda _1 /\lambda _2 } \right)^m }}
{{m!}}\left( {1 - \rho } \right)^{k + 1 - m} } F\left( {\lambda _1
,k + m,\frac{{\lambda _1 /\lambda _2 }}
{{1 - \rho }}} \right) \hfill \\
\end{gathered}
\end{equation}
Similarly, we evaluate $R_k^2 $ by using the complementary
incomplete gamma function $\Gamma \left( {n + 1,x} \right) = n!e^{ -
x} \sum\limits_{m = 0}^n {\frac{{x^m }} {{m!}}} $ [16]. It yields
\begin{equation}
R_k^2  = k!\sum\limits_{m = 0}^k {\frac{{\left( {\lambda _2 /\lambda
_1 } \right)^m }} {{m!}}} \left( {1 - \rho } \right)^{k + 1 - m}
F\left( {\lambda _2 ,k + m,\frac{{1 + \lambda _2 /\lambda _1 }} {{1
- \rho }}} \right)
\end{equation}
\end{IEEEproof}

\subsection{outage probability of secrecy
capacity}The secrecy capacity can also be characterized in terms of
the outage probability for a target secrecy rate. The outage
probability can be calculated according to
\begin{align*}
\begin{gathered}
  P_{out} \left( R \right) = 1 - P_r \left( {C_s \left( {\alpha ,\beta } \right) > R} \right) \hfill \\
   = 1 - P_r \left( {\alpha  > e^R \left( {1 + \beta } \right) - 1} \right) \hfill \\
\end{gathered}
\end{align*}
\begin{theorem}
Similarly invoking again the infinite-series representation of  $I_0
\left( x \right)$,  the outage probability is equivalent to
\begin{equation}
\begin{gathered}
  P_{out} \left( R \right) =1- \exp \left( { - \frac{y}
{{1 - \rho }}} \right)\sum\limits_{k = 0}^\infty  {c_k
k!\sum\limits_{m = 0}^k {\frac{1} {{m!}}} \left( {\frac{\mu }
{{1 - \rho }}} \right)^m }  \hfill \\
   \times \sum\limits_{n = 0}^m {\left( \begin{gathered}
  n \hfill \\
  m \hfill \\
\end{gathered}  \right)} \left( {\frac{y}
{\mu }} \right)^{m - n} \left( {\frac{{1 - \rho }}
{{1 + \mu }}} \right)^{k + n + 1} \Gamma \left( {k + n + 1} \right) \hfill \\
\end{gathered}
\end{equation}
where $y = \left( {e^R  - 1} \right)/\lambda _1 $ and $\mu  = e^R
\lambda _2 /\lambda _1 $. \end{theorem}

At this point in this study, the closed-form expressions of secrecy
capacity have been derived for a correlated Rayleigh fading channel.
In all cases of practical significance, the infinite series
representations can be truncated without sacrificing numerical
accuracy. We also note that the results in [7] are a special case of
our result by assuming independent channels.
\section{Simulation Results}
In Fig. 2, we plot the secrecy capacity versus SNR over correlated
fading channels in the case of the scenario that the SNR of main
channel and eavesdropper's channel are equal.

It is clearly shown that the correlation between the main and the
eavesdropper's channel reduce the secrecy capacity. For comparison,
the limiting value given in [14] also is depicted in Fig 2. The
secrecy capacity converges into the limit of the secrecy capacity
[14] at high SNRs. However, the limiting value is far away form the
secrecy capacity at low and moderate SNRs.

\begin{figure}
\centering
\includegraphics[width=3.0in, height=3in]{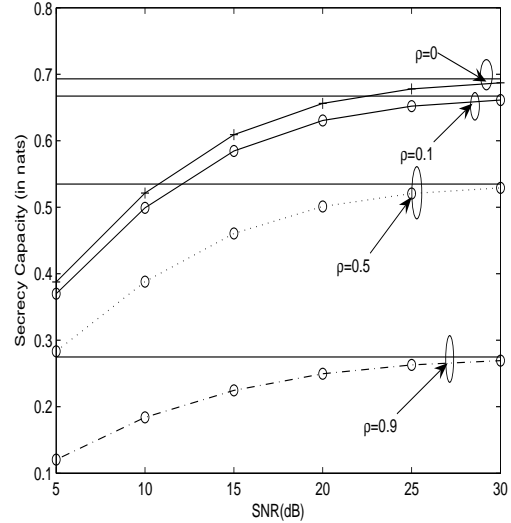}
\vspace{-0.3cm} \caption{The average secrecy capacity versus SNR
over the correlated fading channels. The average SNR of main channel
and eavesdropper's channel are equal. The solid lines indicates the
limiting value given in [14].}
\end{figure}

\section{Conclusion}
In this paper, we investigate the secure communication over
correlated Rayleigh fading Wyner's wiretap channel. The closed form
expressions for average secrecy capacity and outage probability are
derived assuming the full channel state information available.

\section*{Acknowledgment}
This work is supported by National Science Foundation of China under
Grant 60802007 and supported by the Research Fund of National Mobile
Communications Research Laboratory£¬Southeast University 2009A10.

\end{document}